\documentclass{emulateapj}

\slugcomment{Accepted for publication in Astrophysical Journal Letters}
\shorttitle{Clustering of DRGs}
\shortauthors{Quadri et al.}

\begin{document}

\title{A Confirmation of the Strong Clustering of Distant Red Galaxies
  at $2 < z <3$}

\author{
Ryan F.~Quadri\altaffilmark{1},
Rik J.~Williams\altaffilmark{1},
Kyoung-Soo Lee\altaffilmark{2},
Marijn Franx\altaffilmark{1},
Pieter van Dokkum\altaffilmark{3}
Gabriel B.~Brammer\altaffilmark{3}
}

\altaffiltext{1}{Leiden Observatory, Leiden University, NL-2300 RA,
  Leiden, Netherlands}
\email{quadri@strw.leidenuniv.nl}
\altaffiltext{2}{Yale Center for Astronomy and Astrophysics, Yale
  University, New Haven, CT 06520}
\altaffiltext{3}{Department of Astronomy, Yale University, New Haven,
  CT 06520}

\begin{abstract}
  Recent studies have shown that distant red galaxies (DRGs), which
  dominate the high-mass end of the galaxy population at $z \sim 2.5$,
  are more strongly clustered than the population of blue star-forming
  galaxies at similar redshifts.  However these studies have been
  severely hampered by the small sizes of fields having deep
  near-infrared imaging.  Here we use the large UKIDSS Ultra Deep
  Survey to study the clustering of DRGs.  The size and depth of this
  survey allows for an unprecedented measurement of the angular
  clustering of DRGs at $2 < z_{\rm phot} < 3$ and $K<21$.  The
  correlation function shows the expected power law behavior, but with
  an apparent upturn at $\theta \lesssim 10\arcsec$.  We deproject the
  angular clustering to infer the spatial correlation length, finding
  $10.6 \pm 1.6 h^{-1} \textrm{Mpc}$.  We use the halo occupation
  distribution framework to demonstrate that the observed strong
  clustering of DRGs is not consistent with standard models of galaxy
  clustering, confirming previous suggestions that were based on
  smaller samples.  Inaccurate photometric redshifts could
  artificially enhance the observed clustering, however significant
  systematic redshift errors would be required to bring the
  measurements into agreement with the models.  Another possibility is
  that the underlying assumption that galaxies interact with their
  large-scale environment only through halo mass is not valid, and
  that other factors drive the evolution of the oldest, most massive
  galaxies at $z \sim 2$.
\end{abstract}

\section{Introduction}
\label{sec:introduction}

Thus far, the precise measurements of galaxy clustering at $z \gtrsim
2$ that are necessary for meaningful physical interpretation have only
been possible for the relatively blue, star-forming galaxies that
dominate optical surveys \citep{adelberger05, lee06, ouchi05}.
However the most massive galaxies at these redshifts tend to be faint
in the optical, and are more appropriately selected in the
near-infrared \citep[e.g.][]{vandokkum06}.  In particular, galaxies
meeting the $J-K > 2.3$ criterion for distant red galaxies
\citep[DRGs;][]{franx03} have been shown to dominate the high-mass end
of the galaxy mass function \citep{vandokkum06, marchesini07a,
  rudnick06}.  Measurements of DRG clustering have been severely
hampered by small fields and by the near-complete reliance on
photometric redshifts.  Using the largest DRG sample then available,
\citet{quadri07a} confirmed the very strong clustering found by
previous authors \citep{daddi03, grazian06}.  They also showed
that strong clustering implies that DRGs occupy dark matter halos with
$M \gtrsim 10^{13} \rm{M_\odot}$.  However, the number density of
these dark matter halos is at least an order of magnitude smaller than
the number density of DRGs, suggesting that the DRG clustering
measurements are incompatible with models of dark matter clustering.

Here we use a $\sim$0.65$\rm{deg}^2$ field from the UKIRT Infrared
Deep Sky Survey (UKIDSS), which is $\sim$8 times larger than the area
used by \citet{quadri07a} but with similar NIR depth, to determine
whether the strong observed clustering of DRGs found in previous
studies was an artifact due to limited field sizes or whether some
other explanation must be found.

We use $(\Omega_M, \Omega_\Lambda, \sigma_8, h) = (0.3, 0.7, 0.9,
0.7)$.  Small changes in these parameters do not affect our basic
conclusions.  Magnitudes are given in the Vega system, except where
noted.

\section{Data}
\label{sec:data}

The UKIDSS project covers different areas to different depths; here we
make use of the deepest UKIDSS dataset, known as the Ultra Deep Survey
(UDS).  We use the UDS Data Release 1 images \citep{warren07}, which
reach 5 $\sigma$ point-source depths of $J \sim 23$ and $K \sim 21.6$.
We note that \citet{foucaud07} used the UDS Early Data Release to
study the clustering of DRGs down to $K \sim 19$, however such bright
DRGs lie primarily at $z < 2$ and may not be directly relevant to the
galaxies that are the subject of this work.  Most of the UDS field has
coverage in the optical bands from the Subaru-XMM deep Survey
(SXDS)\footnote{http://www.naoj.org/Science/SubaruProject/SXDS/}.  We
use the beta-release of the $BRi'z'$ images, which reach depths of
$\sim$25.3--27.5 (AB magnitudes).  Finally, we combine these data with
$3.6\mu$ and $4.5\mu$ imaging from the \emph{Spitzer} Wide-Area
Infrared Extragalactic Survey \citep[SWIRE;][]{lonsdale03}.  The
procedures used to create a multicolor $K$-selected catalog from these
imaging data are detailed by \citet{williams08}.

\section{Photometric redshifts}
\label{sec:zphots}

Accurate redshift information is fundamentally important for
clustering studies.  Photometric redshifts were calculated using the
EAZY software
\citep{brammer08}\footnote{http://www.astro.yale.edu/eazy/}.  EAZY
fits linear combinations of galaxy templates to the observed
photometry using a Bayesian prior for the distribution of redshifts as
a function of apparent magnitude.  The template set was carefully
chosen to provide high-quality photometric redshifts for $K$-selected
galaxies in several current deep surveys.

To assess the accuracy of our photometric redshifts we compare to a
sample 119 spectroscopic redshifts available in this field \citep[for
further description, see][]{williams08}.  The normalized median
absolute deviation of $\Delta z/(1+z)$ is $\sigma_{\rm NMAD} = 0.033$.
As nearly all spectroscopic redshifts in this field are at $z<1.5$,
and none are for DRGs, we use public data on the well-studied Chandra
Deep Field-South (CDF-S) to obtain an estimate of the expected
photometric redshift accuracy for the galaxies currently under
consideration. We use the photometric catalog of \citet{wuyts08}, and
exclude the $UVH$ filters in order to match the filter set available
in the UDS.  Restricting the sample to the DRGs in CDF-S with
spectroscopic redshifts, we find $\sigma_{\rm NMAD} = 0.065$, with no
apparent dependence on DRG redshift and negligible systematic offsets.
This suggests that the DRG photometric redshifts in our study also
have an accuracy of $\sigma_{\rm NMAD} \sim 0.06-0.07$, although this
can only be taken as a rough indication.

The overall redshift distribution of the sample is used to deproject
the observed angular correlation function in order to infer the
spatial correlation function.  This distribution is often estimated
simply using a histogram of the photometric redshifts, without taking
into account the redshift uncertainties.  Because we select DRGs in a
specific photometric redshift range, it is expected that random errors
will cause galaxies to scatter into the sample from both lower and
higher redshifts.  It is also expected that galaxies within the
redshift selection window will scatter out, however this will not
affect the clustering so long as such galaxies are drawn randomly from
the sample (systematic errors which cause galaxies to scatter into or
out of the selection window are much more problematic).  To estimate
the \emph{intrinsic} redshift distribution of the galaxies in our
sample, we use Monte Carlo simulations in which the observed
photometry is perturbed according the the photometric uncertainties
\citep[see also][]{quadri07a}.  This procedure broadens the
distribution relative to what would be inferred from a histogram of
photometric redshifts, resulting in a higher correlation length; we
return to this point in \S~\ref{sec:discussion}.

\section{Measurements of the DRG  correlation function}
\label{sec:wtheta}

For a detailed description of the techniques used to perform
correlation function measurements, see \citet{quadri07a}.  Here we
briefly describe the method, and point out differences between the
method used here and in the previous paper.  The angular two-point
correlation function is calculated using the \citet{landy93}
estimator.  Uncertainties are estimated using bootstrap resampling,
which yields uncertainties that are significantly larger than those
expected from Poisson statistics.  We follow the method of
\citet{adelberger05} to correct for the integral constraint, although
the correction is small given the large size of the field.

A potentially serious problem for measurements of the auto-correlation
function arises from variable sensitivity or calibration errors across
the images.  We measured the correlation function of stars in order to
test whether such variations induce artificial clustering, obtaining
the expected result that stars are unclustered.  We have also measured
the clustering of DRGs selected from catalogs in which the $J$ and $K$
zeropoints are varied within their uncertainties independently for
each of the pointings that make up the UDS mosaic.  Although it is
difficult to rule out significant errors in the correlation function
arising from such variations, it appears that they are not a dominant
source of error.

The redshift distribution of galaxies selected using the DRG color
criterion is rather wide, extending from $z \sim 1$ to at least $z
\sim 3.5$ \citep[e.g.][]{grazian06,quadri07b}.  We therefore impose
the additional criterion $2 < z_{\rm phot} < 3$.  Over this redshift
range, the DRG criterion effectively selects galaxies with red
rest-frame optical colors \citep{quadri07b}.  We also limit the sample
to objects with $K<21$, ensuring that the catalog is complete and that
objects have a sufficient signal-to-noise ratio to calculate accurate
photometric redshifts.  We visually inspected each object in the
sample, rejecting those that appeared to be artifacts in the images or
were otherwise deemed to have unreliable photometry.  Objects were
also rejected if the templates used for photometric redshifts gave
poor fits.  The final sample consists of 1528 DRGs with $K<21$.

\begin{figure}
  \epsscale{1.1} \plotone{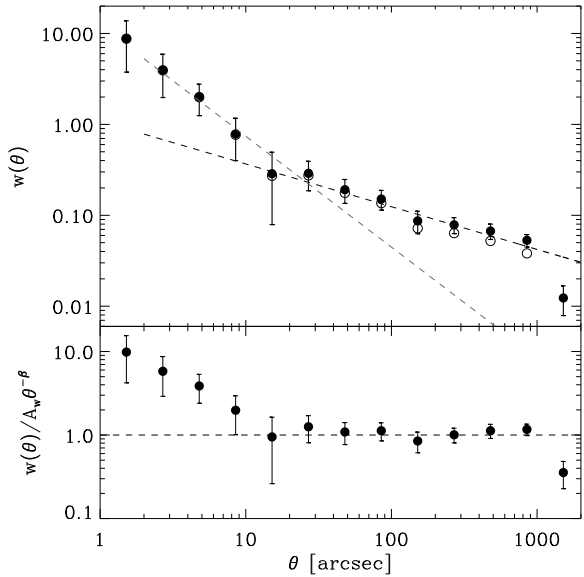}
  \caption{\emph{Top:} The angular correlation function of DRGs with
    $2 < z_{\rm phot} < 3$ and $K<21$.  The open circles illustrate the
    observed values, while the filled circles have been corrected for
    the integral constraint.  Two power-law fits are shown.  The gray
    dashed line indicates the best fitting power law over $2\arcsec <
    \theta <40\arcsec$; it is apparent that this power law does not
    provide a good fit over the entire observed range.  The black
    dashed line indicates the best-fitting power law over $40\arcsec <
    \theta <500\arcsec$.  \emph{Bottom:} The departure of $w(\theta)$
    from the $40\arcsec < \theta <500\arcsec$ power law.}
  \label{fig:acf}
\end{figure}

Figure \ref{fig:acf} shows the angular correlation function
$w(\theta)$ of DRGs.  Our measurements confirm the strong angular
clustering previously reported by \citet{grazian06} and
\citet{quadri07a}, but with reduced uncertainties and extending to
much larger angular scales.  Immediately apparent is the departure of
$w(\theta)$ from a power law on small scales.  To illustrate the
significance of this excess clustering signal, and to show how the
results differ from what would be measured using a smaller field, we
fit a power law $A_w\theta^{-\beta}$ over the range $2\arcsec < \theta
<40\arcsec$ and over $40\arcsec < \theta <500\arcsec$.  The former fit
yields $(A_w,\beta) = (12 \pm 8, 1.2 \pm 0.3)$ while the latter fit
yields $(A_w,\beta) = (1.1 \pm 0.8, 0.47 \pm 0.14)$.  Note that there
is significant covariance between the two fitting parameters, so the
uncertainties should be taken as representative only.  As noted by
previous authors \citep{adelberger05,lee06,quadri07a}, the large-scale
fit is preferred when comparing galaxy samples, as it better probes
the clustering of dark matter halos that host the galaxies.

The bottom panel of Figure \ref{fig:acf} shows the residuals
$w(\theta)$ from the large-scale power law.  A small-scale excess has
also been observed in large samples of optically selected galaxies at
high redshift \citep[e.g.][]{ouchi05, lee06}, and for various galaxy samples
at $0 \lesssim z \lesssim 2$ \citep[e.g.][]{zehavi05}.

The spatial correlation length $r_0$ can be estimated from $w(\theta)$
using the Limber projection; see e.g. \citet{quadri07a} for further
details.  The result is $r_0 = 10.6 \pm 1.6 h^{-1} \textrm{Mpc}$,
where we have used the large-scale fitting parameters and the
uncertainty is estimated using the bootstrap simulations described
above.  This uncertainty accounts for the uncertainty in $\gamma$, but
not in the redshift distribution $N(z)$.  We use the method of
\citet{hamana04} to estimate the linear bias directly from
$w(\theta)$, finding $b=5.0 \pm 0.4$.

\section{The Halo Model}
\label{sec:hod}

\citet{quadri07a} found that DRGs significantly outnumber the dark
matter halos that are clustered strongly enough to host them.  Models
of the halo occupation distribution (HOD) naturally account for this
type of discrepancy by allowing halos to host multiple galaxies, but
\citet{quadri07a} suggest that the large numbers of galaxies that
would be required to share halos can be ruled out by the observed
small-scale clustering.  Given that the DRG clustering results
presented in this work are largely consistent with those from previous
works, it is expected that the observations are still in conflict with
the models.  In this section we use an HOD model to show this
explicitly.

In the HOD framework, the galaxy correlation function is understood to
be the sum of two components.  On large scales the correlation
function follows that of the host dark matter halos.  On small scales
there is an additional contribution from galaxy pairs within
individual halos.  We follow the modeling procedures described by
\citet[]{lee06}, to which we refer the reader for details.  Briefly,
we calculate the number density and bias of halos using the
prescriptions of \citet{sheth99,sheth01}.  The halo occupation number,
which describes the number of galaxies per halo, is parameterized as
$N_{occ}(M_h) = (M_h/M_1)^\alpha$ for halo mass $M_h>M_{min}$ and
$N_{occ}(M_h) = 0$ otherwise.  Thus there are three free parameters,
$M_1$, $M_{min}$, and $\alpha$.  If the number density $n_g$ of
galaxies is known, then one of these parameters can be fixed for
assumed values of the other two parameters.  We obtain a rough
estimate of $n_g \approx 6.5 \times 10^{-4} h^{3} \rm{Mpc}^{-3}$ using
the effective volume probed by our sample, which is determined using
the redshift selection function described in \S~\ref{sec:zphots}.
This density is in reasonable agreement with an independent estimate
of $n_g \approx (5.0 \pm 0.9) \times 10^{-4} h^{3} \rm{Mpc}^{-3}$,
which is based on the luminosity functions of \citet{marchesini07a}.

\begin{figure}
  \epsscale{1.1} 
  \plotone{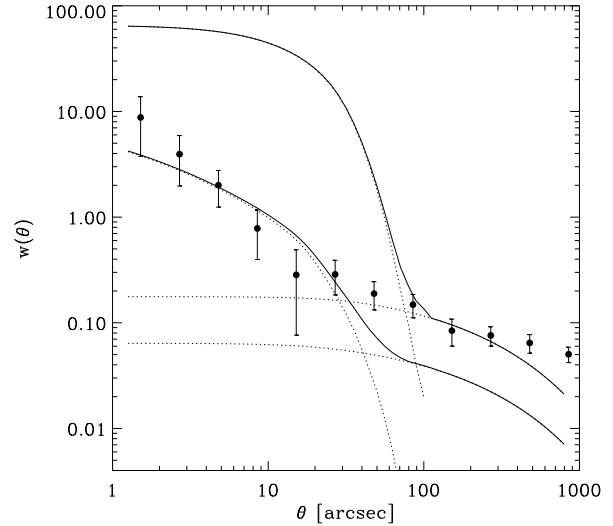}
  \caption{The angular correlation function and the best-fitting HOD
    models.  The solid lines indicate the models, and the dotted lines
    indicate the contributions from galaxy pairs within single halos
    and within separate halos.  The lower model provides the best fit
    over the range $2\arcsec < \theta < 500\arcsec$, with fitting
    parameters $(M_{min},M_1,\alpha) = (1.3 \times 10^{12}
    \rm{M_\odot}, 1.3 \times 10^{13} \rm{M_\odot}, 0.9)$.  The upper
    model provides the best fit over $50\arcsec < \theta <
    500\arcsec$, with fitting parameters $(M_{min},M_1,\alpha) = (1.8
    \times 10^{13} \rm{M_\odot}, 1.3 \times 10^{12} \rm{M_\odot},
    0.9)$.  It is apparent that neither model adequately describes the
    data.}
  \label{fig:hod}
\end{figure}

Figure \ref{fig:hod} shows two models chosen according to $\chi^2$
fits.  The lower solid line shows a model that is fit over $2\arcsec <
\theta < 500\arcsec$.  While this model provides an adequate fit on
smaller angular scales, it systematically under-predicts the
clustering on larger scales.  To better illustrate the nature of the
disagreement, it is useful to inspect the upper solid line, which
shows a model that is fit only over $50\arcsec < \theta < 500\arcsec$.
While this model provides an adequate fit at larger scales, the
small-scale fit is unacceptable.  As already noted, the fundamental
reason that no model can fit the data is that the strong clustering on
large scales implies that DRGs must occupy very massive halos.  But
DRGs outnumber these halos by a factor of $\sim$20, which is only
possible if individual halos host a large number of DRGs.  This would
mean that each DRG has a high probability of having several neighbors
in the immediate vicinity, leading to a very prominent small-scale
excess in $w(\theta)$.\footnote{Specifically, the amplitude of the
  one-halo term depends on the second factorial moment of
  $N_{occ}(M_h)$, which we parameterize following \citet{bullock02}.
  Note that this particular choice does not affect the main result of
  this section because, for the high $\langle N_{occ} \rangle$ value
  found here, it is generically expected that $N_{occ}(M_h)$ follows a
  Poisson distribution for a fixed $M_h$ \citep[e.g.][]{zheng05}; this
  fully specifies the moments.}  As can be seen, the observed excess
is much smaller than expected.

\section{Discussion}
\label{sec:discussion}

We have used the UKIDSS-UDS to perform the first precise measurement
of the clustering of red, $K$-selected galaxies at $2 < z_{\rm phot} <
3$.  These DRGs show strong angular clustering that is well-described
by a power law, but with an excess at small scales.  We use
photometric redshifts to deproject the angular clustering, finding the
spatial correlation length $r_0 = 10.6 \pm 1.6 h^{-1} \textrm{Mpc}$.
This value is comparable to that measured for luminous red galaxies in
the local universe \citep{zehavi05}, however DRGs are significantly
more numerous.  We show that standard models of halo occupation
statistics are unable to simultaneously reproduce the observed
clustering and number density, because DRGs outnumber their inferred
host dark matter halos by too large a margin.

\begin{figure}
  \epsscale{1.1}
  \plotone{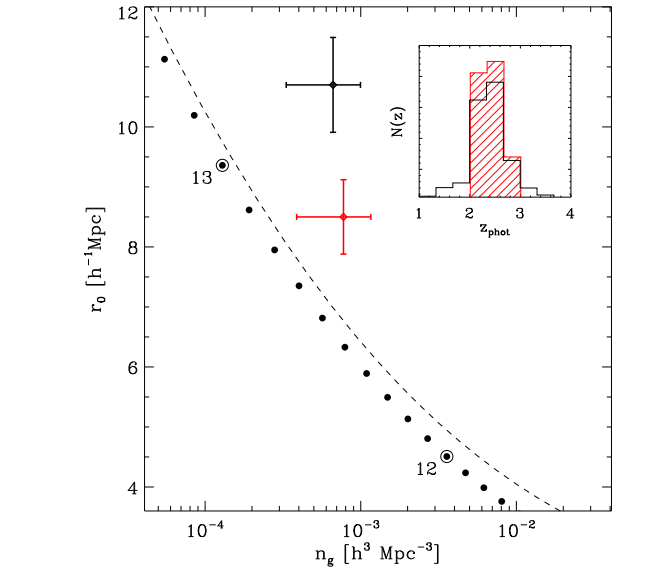}
  \caption{Correlation length versus number density.  The black data
    point shows the values calculated using the preferred redshift
    distribution (\emph{inset:} black histogram), while the red data
    point shows how the results change when using a narrower redshift
    distribution (\emph{inset:} red hatched histogram).  For the
    purposes of this figure we have fixed the power law index of the
    correlation function to $\gamma=1.5$, and assigned conservative
    $50\%$ uncertainties to the observed galaxy number density.  The
    dashed line shows the approximate relation for HOD models of
    mass-selected galaxies based on the simulations of \citet{zheng05}
    and \citet{kravtsov04}.  The filled circles show the approximate
    relation for dark matter halos; the larger open circles are
    labeled with $\log (M_h/\rm{M_\odot})$.  }
  \label{fig:r0_n}
\end{figure}

The most obvious explanation is that we have used the incorrect
redshift distribution in deprojecting the angular correlation
function.  A narrower distribution would reduce the correlation length
(while a moderate shift in the overall distribution makes a relatively
smaller difference).  However, a narrower redshift distribution would
also decrease the effective volume probed by our sample, thereby
increasing $n_g$.  We illustrate these effects in Figure
\ref{fig:r0_n}, which shows the observed $r_0$ and $n_g$ compared to
the range of values for a typical HOD model.  It also shows how
estimating $N(z)$ directly from the unperturbed photometric redshifts
--- which, as mentioned in \S~\ref{sec:zphots}, represents the extreme
assumption of no random photometric redshift errors --- affects the
results.  It may still be the case that we are subject to
\emph{systematic} redshift errors, but we also note that a
significantly narrower $N(z)$ would adversely affect the reasonably
good agreement between our estimate of $n_g$ and the luminosity
functions derived by \citet{marchesini07a}.  Finally, we have verified
that our basic results hold when using a different photometric
redshift code \citep[HYPERZ][]{bolzonella00} and with a different
template set \citep{bruzual03}.

Given the apparently high quality of our photometric redshifts, as
well as the consistency with previous results for the clustering of
DRGs, it is worth considering alternative explanations.  One
possibility is that current HOD models are too simplistic, and that
massive red galaxies occupy halos in unexpected ways.  The fundamental
assumption underlying these models is that galaxy observables depend
on halo mass, and not on the larger-scale environment or on halo
properties such as structure or age.  But environment may play a role,
for instance via its effect on mass accretion rates
\citep{scannapieco03,furlanetto06}.  Additionally, halo clustering
varies with several halo properties, even at fixed mass; this
phenomenon is generically referred to as ``assembly bias''
\citep[e.g.][]{gao07}.  To the extent that these properties affect
galaxy observables, they will also affect galaxy clustering
measurements.  However, we note that current estimates for the
strength of the assembly bias are too small to account for the
observed discrepancies.  As an example, \citet{gao07} have found that,
at $z \sim 2-3$ and in the relevant halo mass range, halos in the
upper 20$\%$ tail of the distribution of halo spins have a $\sim$20$\%$
larger bias than the mean value.  It might then be supposed that
DRGs occupy less massive and more numerous halos with higher spin.
But the increased number density of these low mass halos is
approximately cancelled by the requirement that DRGs can only occupy
20$\%$ of them, so the discrepancy in number densities is unchanged.

The existence of massive red galaxies at $z \gtrsim 2$ was not
predicted by models of galaxy formation, although progress has been
made on this front \citep[e.g.][but see
\citealt{marchesini07b}]{croton06,delucia07}.  The results shown here
suggest that the conflict between models and observations extends to
the relationship between galaxies and dark matter halos.  However,
clustering measurements are sensitive to a number of systematic
effects, so our conclusions remain tentative.  The most obvious source
of error comes from our use of photometric redshifts.  Ongoing
medium-band NIR observations will significantly reduce the photometric
redshift uncertainties \citep{vandokkum08}, and in the longer term
multi-object NIR spectrographs will also improve the situation.  If
future work confirms our results, clustering measurements such as
those presented here may provide a new way to understand the detailed
relationship between galaxy and halo properties.

\acknowledgements

We thank Danilo Marchesini, Qi Guo, Simon White, and Chuck Steidel for
useful discussions, as well as the anonymous referee for a
constructive report.  This work is based data made public by UKIDSS,
SXDS, and SWIRE teams.  R.F.Q. is supported by a NOVA Postdoctoral
Fellowship.  Support from National Science Foundation grant CAREER
AST-0449678 is also gratefully acknowledged.

\end{document}